\documentclass{webofc}
\usepackage[varg]{txfonts}   
\usepackage{bm}
\begin{document}
\title{Interplay between core and corona from small to large systems}
%
%

\author{\firstname{Yuuka} \lastname{Kanakubo}\inst{1}\fnsep\thanks{\email{y-kanakubo-75t@eagle.sophia.ac.jp}} \and
        \firstname{Yasuki} \lastname{Tachibana}\inst{2}\fnsep\thanks{\email{ytachibana@aiu.ac.jp}} \and
        \firstname{Tetsufumi} \lastname{Hirano}\inst{1}\fnsep\thanks{\email{hirano@sophia.ac.jp}}
}

\institute{Department of Physics, Sophia University, Tokyo 102-8554, Japan
\and
          Akita International University, Yuwa, Akita-city 010-1292, Japan
          }

\abstract{
We present new results in {\it{p}}+{\it{p}} and Pb+Pb collisions at the LHC energies from the updated dynamical core--corona initialization framework (DCCI2).
The fractions of final hadron yields originating from equilibrated and non-equilibrated components are extracted as functions of multiplicity. We find that the contributions from non-equilibrated components are non-negligible even in Pb+Pb collisions and affect $p_T$-integrated multi-particle correlations.
These suggest the importance of non-equilibrated components for the sophisticated extraction of properties of the quark gluon plasma from comparisons between dynamical frameworks and experimental data.
}
\maketitle
\section{Introduction}
\label{intro}
The properties of the quark-gluon plasma (QGP) have been investigated through dynamical frameworks based on relativistic hydrodynamics.
Recently, several attempts from dynamical frameworks are reported \cite{Novak:2013bqa,Sangaline:2015isa,Bernhard:2015hxa,Bernhard:2016tnd,Auvinen:2017fjw,Bernhard:2019bmu,JETSCAPE:2020shq,Nijs:2020ors,Auvinen:2020mpc,Parkkila:2021tqq} to quantitatively extract the QGP transport coefficients from the data-to-model comparison.
This indicates that sophisticated description with dynamical frameworks is needed for a further investigation of transport coefficients.
On the other hand, the possibility of the QGP formation in small colliding systems has been discussed following the report of strangeness enhancement in {\it{p}}+{\it{p}} collisions by the ALICE Collaboration \cite{ALICE:2017jyt}.
Thus, the extension of the applicability of those dynamical frameworks to small colliding systems is indispensable. 

Motivated by these backgrounds,
we build the dynamical core--corona initialization as a state-of-the-art dynamical framework that is capable of describing from small to large colliding systems \cite{Kanakubo:2018vkl,Kanakubo:2019ogh,Kanakubo:2021qcw}.
Under the core--corona picture \cite{Werner:2007bf}, the system generated in high-energy nuclear collisions is described with two components: equilibrated matter (core) and non-equilibrated matter (corona).
In our framework, we realize the dynamical description of the separation of the system into the core and corona by incorporating the picture with the novel dynamical initialization approach \cite{Okai:2017ofp}.

\section{Dynamical Core--Corona Initialization framework}
\label{sec-1}
Under the dynamical initialization approach, 
initial conditions of QGP fluids are generated dynamically via source terms of 
relativistic hydrodynamic equations.
We assume that the system is described with the two components, QGP fluids as core and non-equilibrated partons as corona, respectively,
and consider the energy and momentum conservation for the entire system generated in a collision.
Thus, the depositions of energy and momentum from non-equilibrated partons become sources of QGP fluids,
and which makes it possible to model the generation of initial conditions of QGP fluids with keeping the energy and momentum conservation of the entire system.
The dynamical realization of the core--corona picture is attained  
by modeling the energy and momentum deposition rates of non-equilibrated partons.
Based on the core--corona picture,
non-equilibrated partons with low transverse momentum, $p_T$, and in high-density regions are likely to deposit their energy and momentum while those with high $p_T$ and in low-density regions are not.

A brief summary of the latest version of our framework \cite{Kanakubo:2021qcw,Kanakubo:2022ual} is as follows:
Initial conditions are phase-space distributions of partons after the final-state radiation
obtained from \textsc{Pythia}8 \cite{Sjostrand:2007gs}
and \textsc{Pythia}8 Angantyr \cite{Bierlich:2016smv,Bierlich:2018xfw} for {\it{p}}+{\it{p}} and A+A collisions, respectively.
The dynamical initialization is performed with those initially produced partons
and the system is dynamically separated into the core and corona.
The space-time evolution of the core, QGP fluids, is described with (3+1)-D ideal hydrodynamics \cite{Tachibana:2014lja}
by incorporating the equation of state $s$95$p$-v.1.1 \cite{Huovinen:2009yb}.
Once temperatures of fluids become lower than a switching temperature,
fluids are particlized with \textsc{iS3D} \cite{McNelis:2019auj}, a Monte-Carlo sampler of hadrons from hydrodynamic fields.
The non-equilibrium partons are hadronized via the Lund string fragmentation with \textsc{Pythia}8.
Finally, those directly emitted hadrons are handed to \textsc{Jam} \cite{Nara:1999dz}, a hadronic transport model that performs hadronic rescatterings and resonance decays.

\section{Results}
\label{results}
\begin{figure}
\centering
\sidecaption
\includegraphics[bb=0 0 641 451, width=0.525\textwidth]{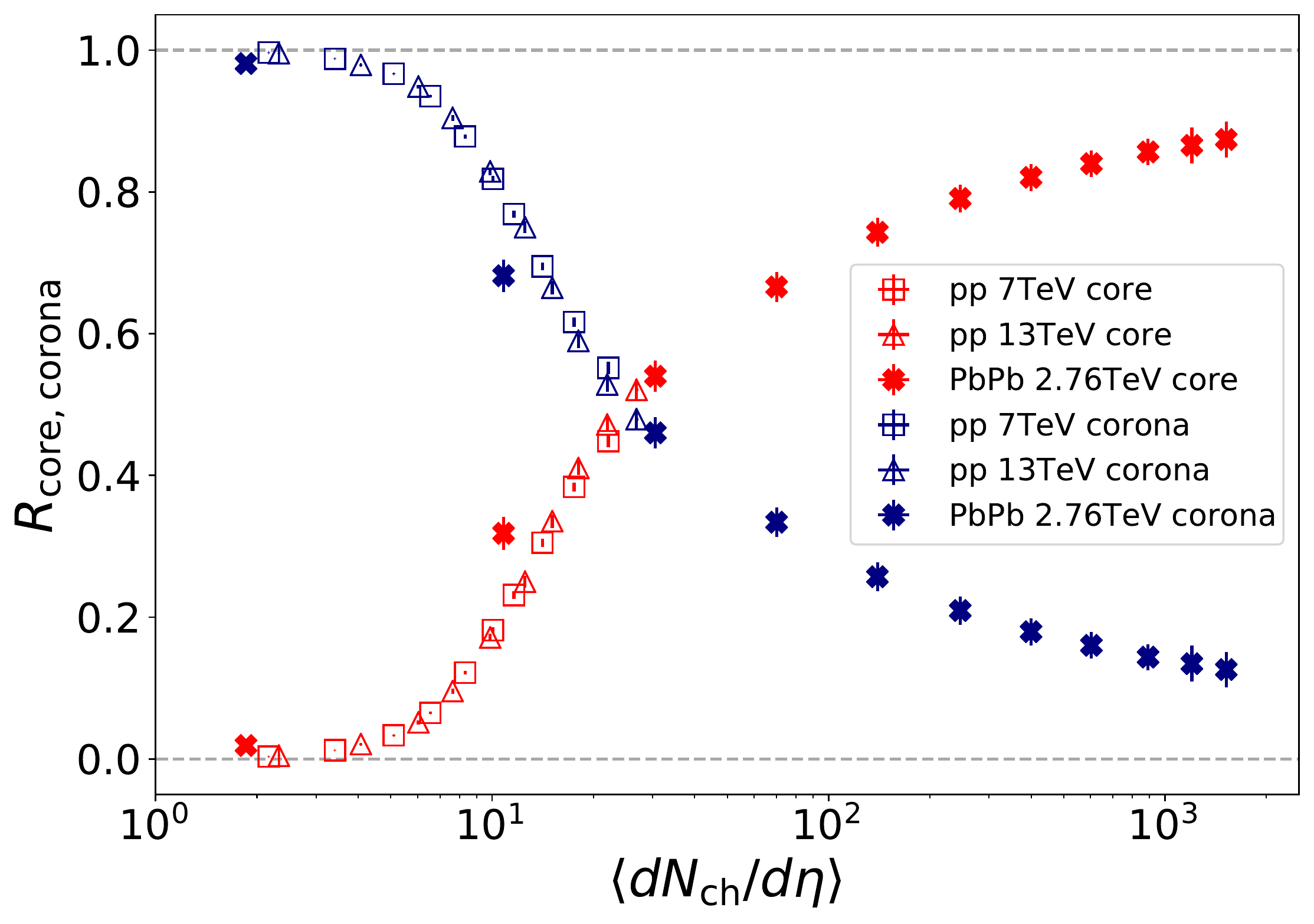}
\caption{Fractions of hadron yields originating from the core and corona as functions of the number of charged particles produced at midrapidity $|\eta|<0.5$.
Results in {\it{p}}+{\it{p}} collisions at $\sqrt{s}=$ 7 and 13 TeV (open squares and triangles, respectively) and Pb+Pb collisions at $\sqrt{s_{\mathrm{NN}}}=2.76$ TeV (closed circles) are shown.
}
\label{fig-1}       
\end{figure}

We simulate {\it{p}}+{\it{p}} and Pb+Pb collisions at the LHC energies with our framework.
Some major parameters control the fraction of generated core and corona components.
We determine them so that the full simulation results can reasonably describe the multiplicity dependence of the yield ratios of omega baryons to charged pions in {\it{p}}+{\it{p}} and Pb+Pb collisions \cite{Kanakubo:2021qcw} \footnote{
Note that some parameters are minor updated from Ref.~\cite{Kanakubo:2021qcw}. For details, see Ref.~\cite{Kanakubo:2022ual}.
}
reported by the ALICE Collaboration 
\cite{ALICE:2017jyt}.
By fixing those parameters, we extract the fractions of hadron yields originating from core and corona as functions of multiplicity near mid-pseudo-rapidity ($|\eta|<0.5$) in {\it{p}}+{\it{p}} collisions at $\sqrt{s}=$ 7 and 13 TeV and Pb+Pb collisions at $\sqrt{s_{\mathrm{NN}}}=2.76$ TeV as shown in Fig.~\ref{fig-1}.
One sees that the fractions of core and corona components show clear multiplicity scaling regardless of the system size or collision energies.
Thus, within our framework, final hadron yields can be one of the reasonable estimators to evaluate the fraction of equilibrated and non-equilibrated components.
We also find that the fraction of core components becomes dominant when multiplicity exceeds $\left \langle dN_{\mathrm{ch}}/d\eta \right\rangle \approx 20$.
It should be also noted that the fraction of core (corona) components does not reach $\approx100\%$ ($\approx0\%$) even at the highest multiplicity bin in Pb+Pb collisions.

\begin{figure*}
\centering
\includegraphics[bb=0 0 628 538, width=0.45\textwidth]{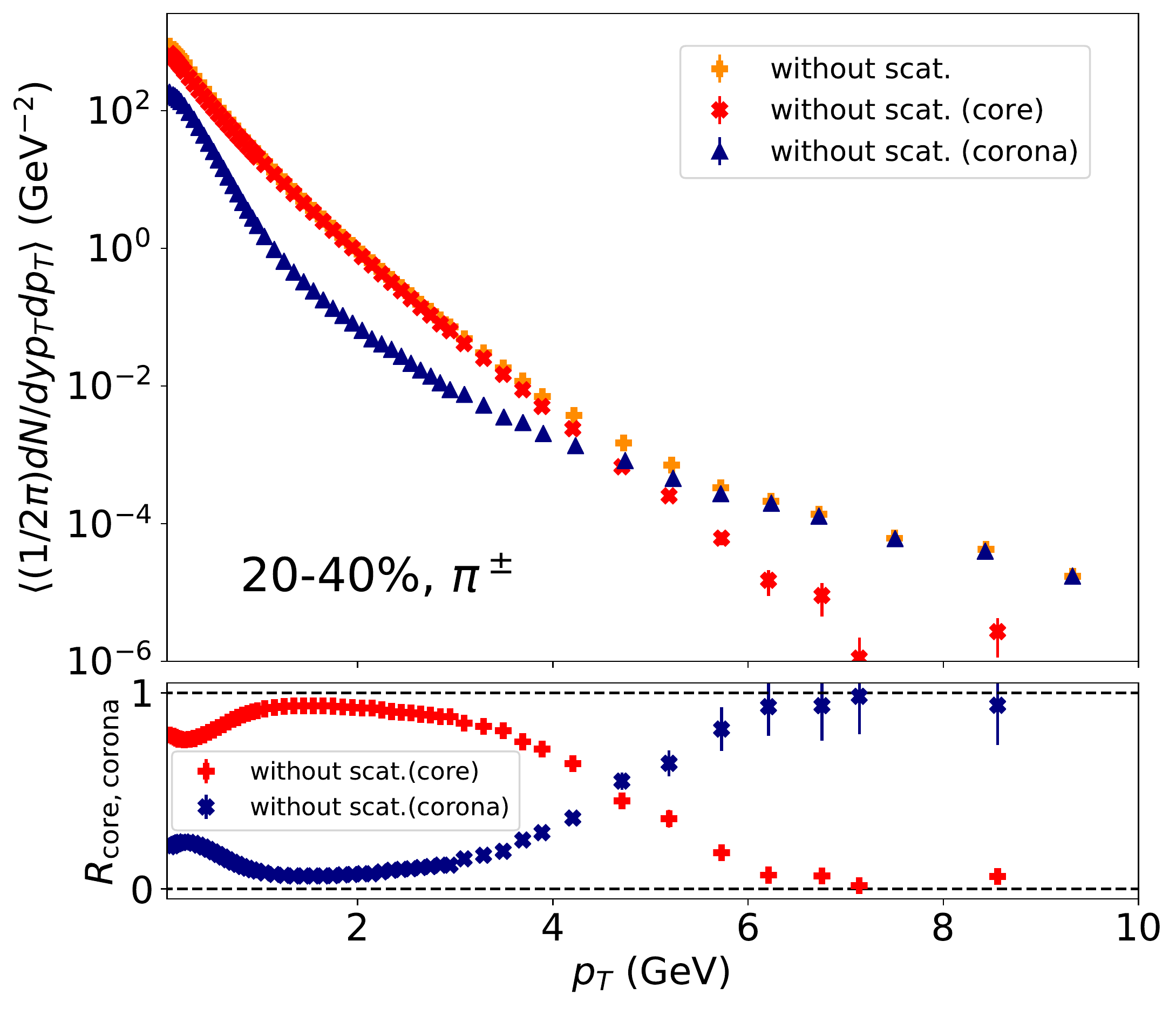}
\includegraphics[bb=0 0 639 458, width=0.50\textwidth]{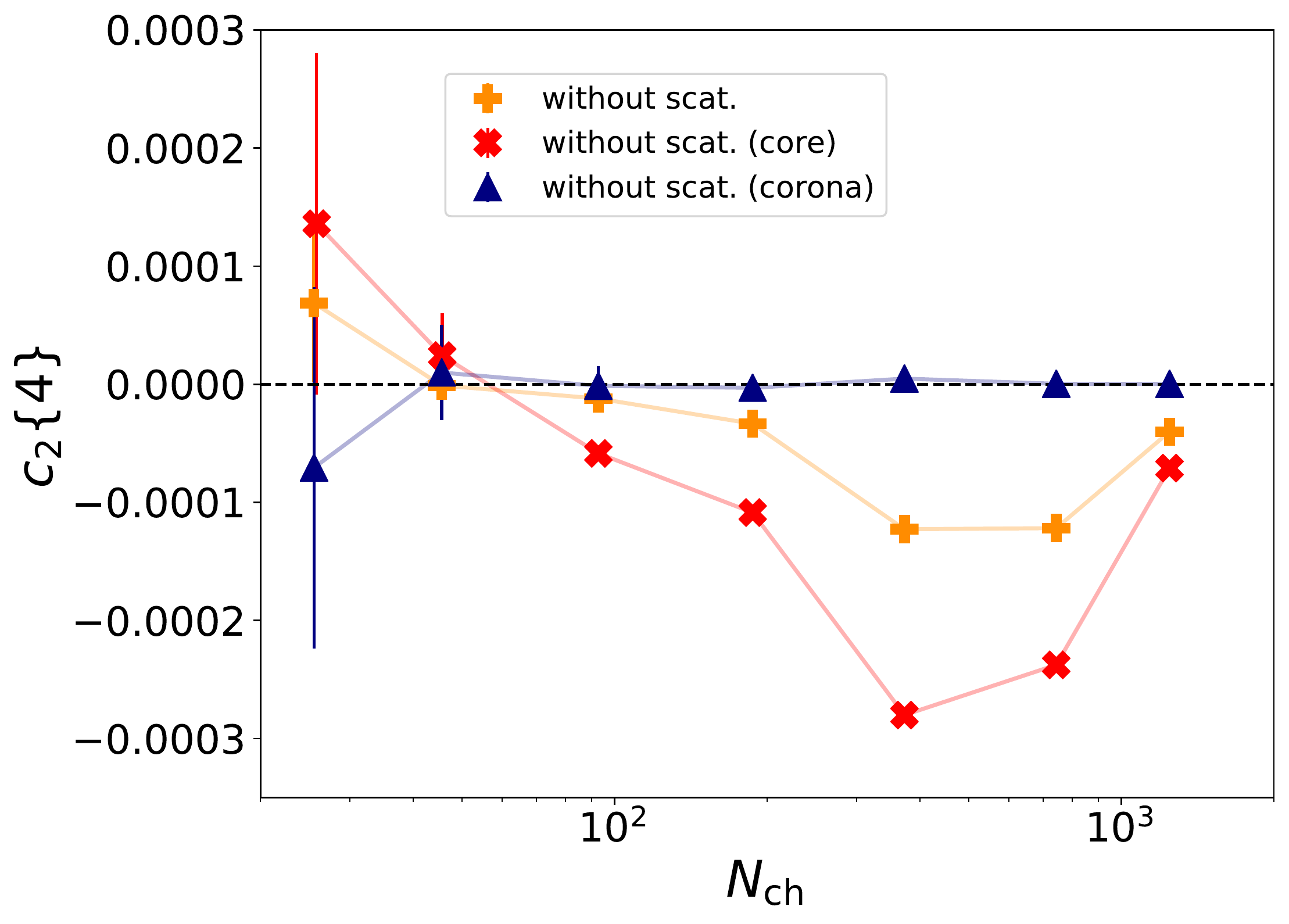}
\caption{(Left) Charged pion $p_T$ spectra obtained from 20-40\% centrality class of Pb+Pb collisions at $\sqrt{s_{\mathrm{NN}}}=2.76$ TeV are shown in the upper panel.
Fractions of core and corona contributions in each $p_T$ bin are shown in the lower panel. 
(Right) Four-particle cumulant of charged particles as a function of the number of charged particles produced at midrapidity obtained in Pb+Pb collisions at $\sqrt{s_{\mathrm{NN}}}=2.76$ TeV.
The solid lines are guides to the eyes.
In both left and right figures, results with
switching off hadronic rescatterings are shown in orange crosses and their breakdowns into the core and corona contributions are shown in red diagonal crosses and blue triangles, respectively. 
}
\label{fig-2}       
\end{figure*}

For a further investigation, we show $p_T$ spectra of charged pions obtained in 20-40\% centrality Pb+Pb collisions at $\sqrt{s_{\mathrm{NN}}}=2.76$ TeV in Fig.~\ref{fig-2} (left).
As an overall tendency, 
the core (corona) component dominates low (high) $p_T$ regions, which is exactly the consequence of implementing the core--corona picture.
However, against the naive expectation of the above tendency, there is a slight enhancement of the corona components at the very low $p_T$ regions ($p_T<1$ GeV).
This can be intuitively understood just as a result of the interplay between the two different shapes of $p_T$ spectra: exponential for the core components and power law for the corona components.
The hadronic productions at such low $p_T$ regions in the $p_T$ spectra of corona components originate from the feed-down from string fragmentation of non-equilibrated partons that lose their initial energy and momentum in dynamical initialization \cite{Kanakubo:2021qcw}.

To see the influence of the existence of the corona components in the very low $p_T$ on observables, 
four-particle cumulant of charged particles, $c_2\{4\}$, as a function of the number of charged particles, $N_{\mathrm{ch}}$, produced near midrapidity obtained from Pb+Pb collisions at $\sqrt{s_{\mathrm{NN}}}=2.76$ TeV is shown in Fig.~\ref{fig-2} (right).
For the kinematic ranges, $c_2\{4\}$ and $N_{\mathrm{ch}}$
are obtained from charged particles with
$|\eta|<0.8$ and $0.2 < p_T < 3.0$ GeV which is the same ranges used in Ref.~\cite{Acharya:2019vdf}.
The results from simulations switching off hadronic rescatterings and from the core components show the negative values of $c_2\{4\}$, while those from the corona components are zero-consistent in the entire range of $N_{\mathrm{ch}}$.
Notably, from the comparisons among those three results, one sees that the absolute values of $c_2\{4\}$ solely from the core components are diluted due to the existence of the corona contributions.

\section{Summary}
We reported new results from the updated version of the dynamical core--corona initialization framework.
We found that the fractions of the core and corona components show clear scaling with multiplicity regardless of the system size or collision energies and that the core components become dominant above $\langle dN_{\mathrm{ch}}/d\eta \rangle \approx 20$.
We also showed that there are non-negligible contributions from corona components, which significantly affect $c_2\{4\}$ as a function of $N_{\mathrm{ch}}$.
Our results suggest the importance of the non-equilibrium components even in heavy-ion collisions to extract the QGP properties quantitatively.

\bibliography{SQM2022-templates/template}

\end{document}